\documentclass[aps,prl,reprint,groupedaddress]{revtex4-2}
\usepackage{amsfonts}
\usepackage{amssymb}
\usepackage{amsmath}
\usepackage{bbm}
\usepackage{bm}
\usepackage{relsize}
\usepackage{graphicx}

\DeclareMathAlphabet\mathbfcal{OMS}{cmsy}{b}{n}

\newcommand{\be}{\begin{equation}}
\newcommand{\ee}{\end{equation}}
\newcommand{\bea}{\begin{eqnarray}}
\newcommand{\eea}{\end{eqnarray}}
\newcommand{\Eq}[1]{Eq.\,(\ref{#1})}
\newcommand{\Eqs}[2]{Eqs.\,(\ref{#1}) and (\ref{#2})}

\newcommand{\Fig}[1]{Fig.\,\ref{#1}}

\newcommand\wI{\hat{\mathbb I}}
\newcommand\wG{\hat{\mathbb G}}

\newcommand\wP{\hat{\mathbb P}}

\newcommand\wD{\hat{\mathbb D}}

\newcommand\wF{\vec{\mathbb F}}

\newcommand\wA{\vec{\mathbb A}}
\newcommand\wB{\vec{\mathbb B}}

\newcommand{\heps}{\hat{{\pmb{\varepsilon}}}}
\newcommand{\hmu}{\hat{{\pmb{\mu}}}}
\newcommand{\heta}{\hat{{\pmb{\eta}}}}
\renewcommand{\r}{\textbf{r}}

\newcommand{\E}{\textbf{E}}

\newcommand{\D}{\textbf{D}}
\renewcommand{\H}{\textbf{H}}
\newcommand{\B}{\textbf{B}}

\newcommand{\rmfor}{\quad{\rm for}\quad}

\newcommand{\sumint}[1]{\mathrel{\sum_{#1}\!\!\!\!\!\!\!\!\int}}

\begin{document}

\title{Rigorous theory of coupled resonators}

\author{E. A. Muljarov}
\affiliation{%
School of Physics and Astronomy, Cardiff University, Cardiff CF24 3AA, United Kingdom}
\date{\today}

\begin{abstract}

We demonstrate the general failure of the famous concept of tight binding and mode hybridization underlying modern theories of coupled open resonators. In spite of sophisticated examples in the literature, successfully illustrating these theories, the latter fail to describe any planar systems. This includes the simplest possible case of two dielectric slabs placed next to each other or separated by a distance, which is straightforward for verification, due to its analytical solvability. We present a rigorous theory capable of calculating correctly the eigenmodes of arbitrary three-dimensional dispersive coupled resonators in terms of their individual modes, providing insight into the proper mode hybridization and formation of bonding and antibonding supermodes. Planar optical resonators, such as coupled slabs and Bragg-mirror microcavities, are used for illustrative purposes as they allow precise and reliable verification of the theory.

\end{abstract}
\maketitle

\begin{figure}[t]
\includegraphics*[clip,width=0.6\textwidth]{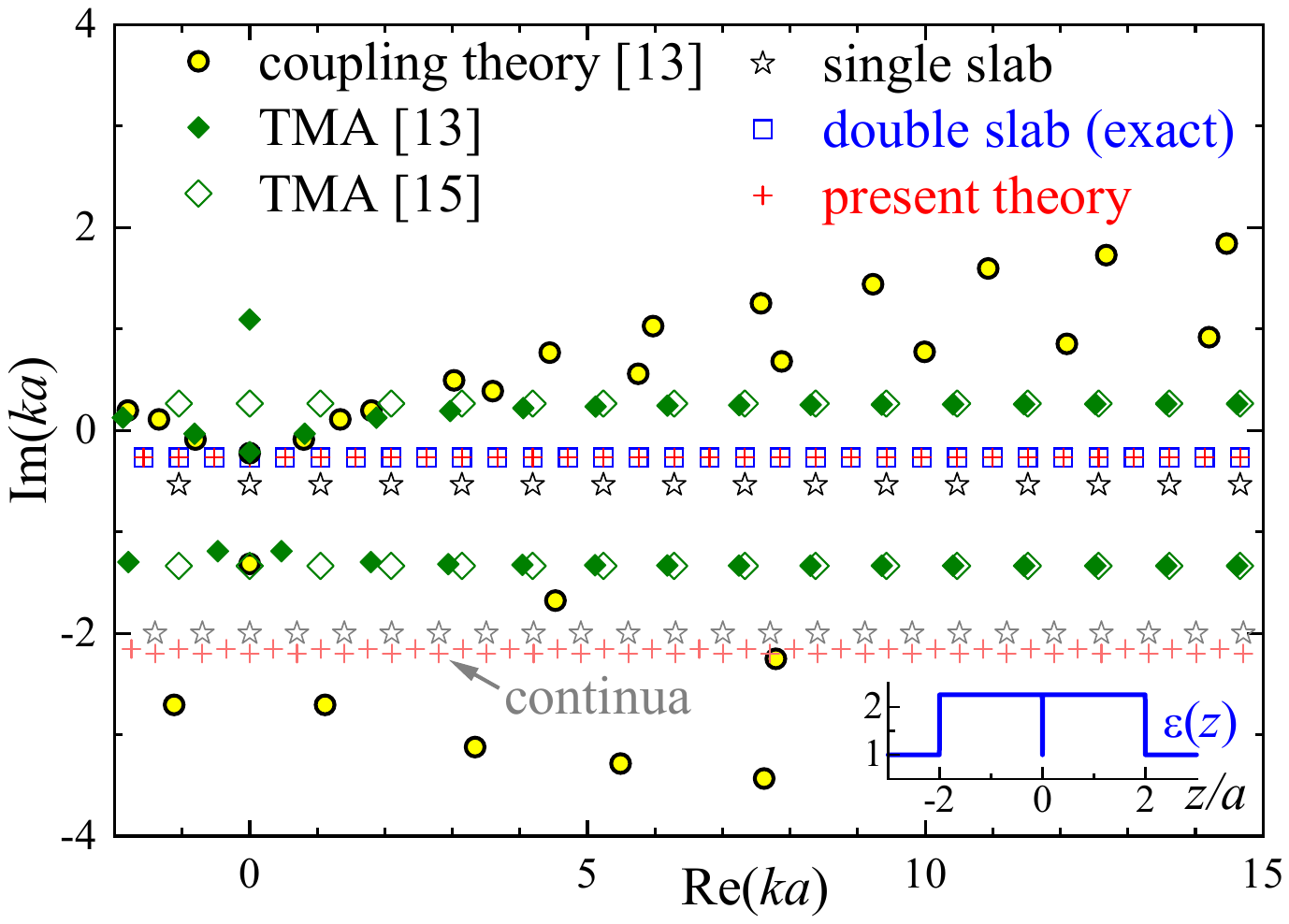}
\vspace{-20.0mm}
		\caption{
RS wave numbers of two identical slabs of width $2a$ and permittivity $\epsilon$ next to each other ($d=0$), calculated exactly (blue open squares), using the coupling theory \cite{LiuPRB20} with a sufficient number of basis RSs ($N=3200$) to reach a visual convergence (circles), TMAs based on Refs.\,\cite{LiuPRB20} and \cite{HughesPRX20} (full and open diamonds), and the present theory (red crosses) using $N=100$ basis RSs (black stars) and 150 discretized modes of the continuum. The latter are shown for the unperturbed (single-) and perturbed (double-slab) systems by gray stars and light-red crosses, respectively. Inset: permittivity profile of the coupled resonators.
 }
\label{Fig:theories}
	\end{figure}

{\em Introduction}. Any resonator is characterized by its eigenmodes, which can be found rather precisely by various analytical or numerical methods \cite{LalanneJOSAA19,WuFP24}. When two or more open resonators are located next to each other or separated by some distance, the modes of each individual resonator are perturbed, and mixing or hybridization of the original eigenstates is expected and in fact observed experimentally, e.g. in photonic molecules \cite{BayerPRL98,MukaiyamaPRL99,AshiliOE06,PreuOE08, RakovichLPR10,BenyoucefOL11,PengOL12,FlattenLPR16,LiLPR17}. It is therefore natural to ask a question: How these hybrid modes of the coupled resonators can be found by using the information about the individual resonators, in particular, their modes? To address this question, several different approaches to finding the modes of coupled optical resonators have been recently developed \cite{VialJO16,LiuPRB20,CogneePhD20,HughesPRX20,BachelardOE22}.  However, in spite of the claims that they are rigorous \cite{LiuPRB20} or accurate enough \cite{HughesPRX20}, supported by various illustrations \cite{VialJO16,LiuPRB20,CogneePhD20,HughesPRX20,BachelardOE22}, none of them work even approximately for planar coupled resonators. This is demonstrated in \Fig{Fig:theories} for the simplest analytically solvable system -- a homogeneous dielectric slab consisting of two identical glass slabs placed next to each other.

The modes of an electromagnetic system, also known in the literature as quasi-normal modes or resonant states (RSs), have a simple analytical form in the case of a dielectric slab surrounded by vacuum, with the mode wave numbers given by
\be
k_n=\frac{1}{2a\sqrt{\epsilon}}\left(\pi n -i \ln \frac{\sqrt{\epsilon}+1}{\sqrt{\epsilon}-1}\right)\,,
\label{slab}
\ee
where $n$ is an integer, $2a$ is the slab thickness, and $\epsilon$ is its permittivity. These wave numbers are the eigenvalues of Maxwell's equations solved with the boundary conditions of outgoing waves normal to the slab \cite{MuljarovEPL10}.
They are shown in \Fig{Fig:theories} for single (black stars) and double slab (blue squares), with the permittivity profile of the latter provided in the inset.
The result of the coupling theory \cite{LiuPRB20}, using the full sets of the RSs of each resonator, are also shown (circles), along with a two-mode approximation (TMA) based on \cite{LiuPRB20} and the TMA by Ren {\em et al.} \cite{HughesPRX20} (open and full diamonds, respectively). Apart from the central ($n=0$) mode, they are all very different from the exact values \Eq{slab}, about a half of them having a positive imaginary part, which is unphysical. One could argue that the TMAs fail here due to the too low quality factor of the modes or too short distance between resonators. This is not the case, however, as we show below: These theories are not applicable also to high-quality (high-Q) modes of coupled Bragg-mirror microcavities (MCs). Furthermore, the famous concept of tight binding turns out to be inapplicable to open resonators, as we show in this Letter below.


The fundamental reason why the theories \cite{VialJO16,LiuPRB20,CogneePhD20,HughesPRX20,BachelardOE22} do not work in the simplest example in \Fig{Fig:theories} is that the RSs, which are complete within the volume of a resonator, are {\it incomplete} outside it, which prevents one from developing any valid expansion needed to properly describe the coupling between resonators. Recent proposals to fill in this gap by RS regularization \cite{AbdelrahmanOSA18,FrankePRL19,KristensenAOP20} were not successful in developing  suitable expansions for coupled resonators \cite{FrankePRA23,LalanneLPR25}, in the best scenario ending up with a highly nonlinear eigenvalue problem \cite{KristensenAOP20}, lacking convergence.  The completeness outside the resonator has been achieved \cite{YanPRB18,SauvanOE22} by supplementing the physical modes (i.e., the RSs) with a large (ideally, infinite) set of unphysical numerical modes which appear in the spectrum due to discretization of differential operators and the use of perfectly matched layers, or with a set of virtual gap modes \cite{SztranyovszkyPRR25} naturally generated by the resonant-state expansion (RSE) \cite{MuljarovEPL10}. However, using this completeness in the context of coupled resonators, would require an excessively large computational domain \cite{footnote1} including the resonators and would not provide any insight into their coupling or hybridization of their modes.


In this Letter, we develop a rigorous theory of coupled resonators, which allows us to calculate their hybridized modes numerically exactly in terms of the modes of the individual resonators. Technically, this is achieved by generalizing the  Mittag-Leffler (ML) expansion of the dyadic Green's function (GF) and extending its validity beyond the resonator boundary. The general theory is developed for arbitrary three-dimensional (3D) dispersive resonators, which can also be magnetic or even bi-anisotropic. However, for clarity of demonstration and verification, illustrations are provided for planar dielectric non-dispersive systems, such as two slabs and two MCs separated by a distance.
Increasing the distance between resonators, the exponential growth of the RSs imposes serious limitations on the applicability of the theory in a form of its poorer or no convergence. This challenge has been successfully addressed by combining the present theory with the RSE which allows us also to rigorously prove the developed formalism.


{\em Two coupled dispersive resonators}. Using the notations introduced in Ref.\,\cite{MuljarovOL18}, we write Maxwell's equations
\be
\nabla \times \E=ik\B\,, \quad  \nabla \times \H=-ik\D
\label{MEcurl}
\ee
for a monochromatic electromagnetic field with a harmonic time dependence $e^{-i\omega t}$ as 
\be
[k\wP(\r;k)-\wD(\r)] \wF(\r)=0\,,
\label{ME}
\ee
where $k=\omega/c$ is the light wave number, with $\omega$ being the light frequency and $c$ the speed of light in vacuum,
$$
\wF(\r)= \left(\begin{array}{c}
\E(\r)\\
i\H(\r)\\
\end{array}\right)
$$
is a 6-dimensional vector comprising  the electric field $\E$ and magnetic field $\H$, and
$\wP(\r;k)$ and $\wD(\r)$ are, respectively, the generalized dispersive permittivity tensor and the curl operator, defined as
\be
\wP(\r;k)=\left(\begin{array}{cc}
\heps(\r;k)&\heta(\r;k)\\
\heta^{\rm T}(\r;k)&\hmu(\r;k)\\
\end{array}\right)\,,
\ \ \ \
\wD(\r)= \left(\begin{array}{cc}
0&\nabla\times\\
\nabla\times&0\\
\end{array}\right)\,.
\label{perm}
\ee
Here $\heps(\r;k)$ and $\hmu(\r;k)$ are respectively, the standard frequency-dependent $3\times3$ permittivity and permeability tensors, $\heta(\r;k)$ is the bi-anisotropy tensor, and T denotes matrix transposition. While we assume reciprocity of the generalized permittivity, implying also $\heps^{\rm T}=\heps$ and $\hmu^{\rm T}=\hmu$, and for illustration use  achiral ($\heta=0$) and nonmagnetic systems, generalizations of the results presented below for non-reciprocal systems \cite{SauvanOE22} are straightforward.


Let us consider two resonators, described by the generalized permittivies $\wP_1(\r;k)$ and $\wP_2(\r;k)$,
and occupying volumes $V_1$ and $V_2$, respectively, which for clarity of derivation are assumed to be not overlapping \cite{footnote2}. The RSs of the full system, comprising both resonators, are then described by \Eq{ME}, in which
\be
\wP(\r;k)= \wP_1(\r;k)+\wP_2(\r;k)-\wP_b\,,
\ee
with $\wP_b$ being a constant tensor of the background generalized permittivity (in case of the vacuum background, $\wP_b=\wI$, where $\wI$ is the $6\times6$ identity matrix), and $k$ and $\wF(\r)$ are, respectively, the wave number and the electromagnetic field of a RS of the full system. The RSs of each resonator are the eigen solutions of Maxwell's equations
\be
[k_n^{(j)}\wP_j(\r;k_n^{(j)})-\wD(\r)] \wF_n^{(j)}(\r)=0\,,
\label{MEi}
\ee
solved with outgoing boundary conditions, where the index $j=1,2$ labels the resonators and $n$ labels the RSs of each resonator. Using the dyadic GF $\wG_j(\r,\r';k)$ of each subsystem, satisfying the inhomogeneous Maxwell's equations,
\be
[k\wP_j(\r;k)-\wD(\r)] \wG_j(\r,\r';k)=\wI\delta(\r-\r')\,,
\label{GF-equi}
\ee
where $\delta(\r-\r')$ is the Dirac delta function in 3D, the formal solution of \Eq{ME} for the full system
can be written as
\be
\wF(\r)=-k\int_{V_2} d\r' \wG_1(\r,\r';k)[\wP_2(\r';k)-\wP_b] \wF(\r')
\label{F1}
\ee
for $\r\in V_1$ and
\be
\wF(\r)=-k\int_{V_1} d\r' \wG_2(\r,\r';k)[\wP_1(\r';k)-\wP_b] \wF(\r')
\label{F2}
\ee
for $\r\in V_2$.

\begin{figure}[t]
\includegraphics*[clip,width=0.4\textwidth]{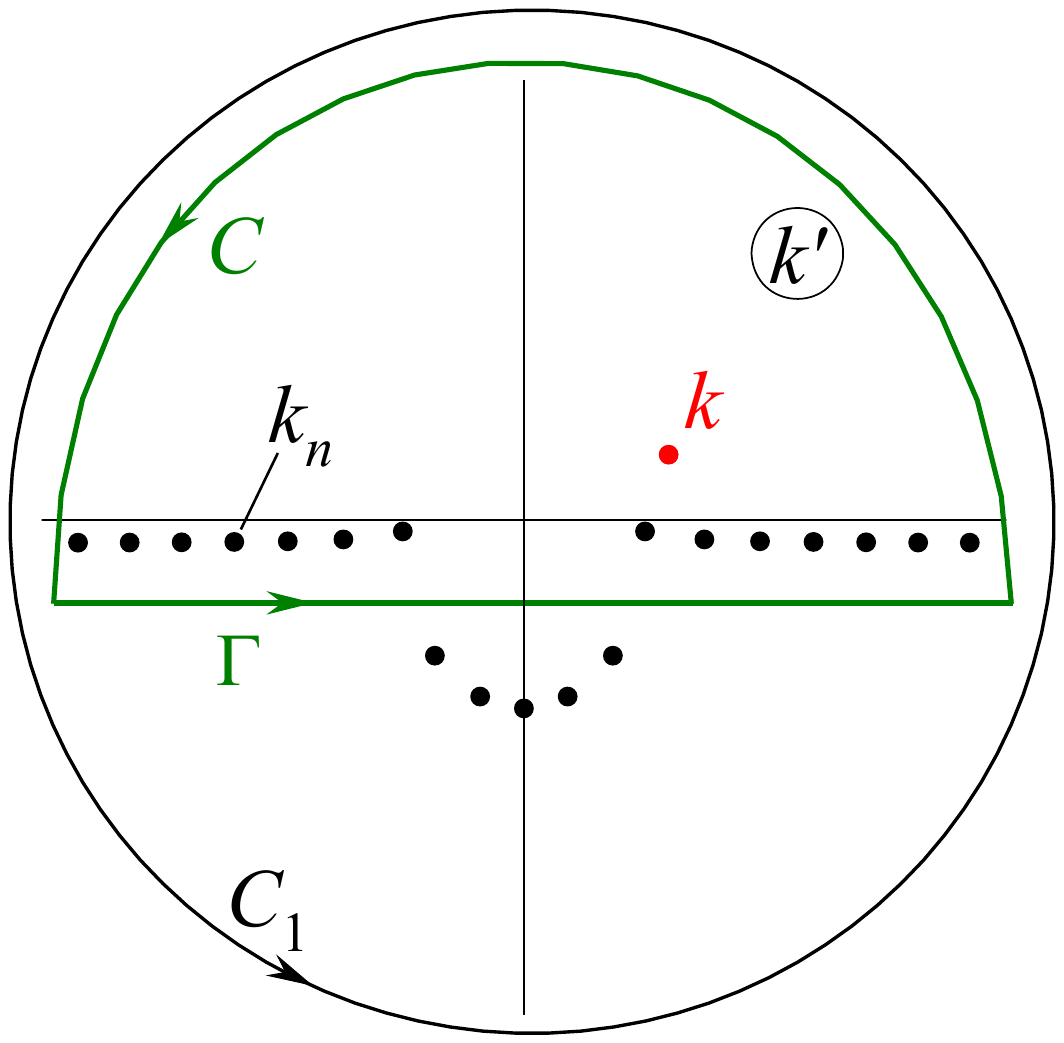}
\vspace{-9.0mm}
\caption{ Full-circle closed contour $C_1$, semicircle contour $C$, and straight-line contour $\Gamma$ in the complex $k'$ plane, along with the poles $k=k_n$ of the GF of a resonator and an additional pole at $k'=k$ (red).
 }
\label{Fig:contour}
	\end{figure}

To find the wave number $k$ and field $\wF(\r)$ of a RS of the full system, using the RSs of each resonator,
one needs to expand the GFs $\wG_j(\r,\r';k)$ in terms of such states. The expansion is known as the ML series of the GF and takes the form \cite{BangNPA78,MuljarovEPL10,DoostPRA13,MuljarovOL18,SI}
\be
\wG_j(\r,\r';k)=\sum_n \frac{\wF_n^{(j)}(\r)\otimes \wF_n^{(j)}(\r')}{k-k_n^{(j)}}\rmfor \r,\r' \in V_j\,,
\label{GF-ML}
\ee
where $\otimes$ denotes the dyadic product of vectors. The ML series \Eq{GF-ML} converges to the correct GF if both coordinates $\r$ and $\r'$ are within the system volume $V_j$, as indicated. This follows from the fact that
\be
\lim_{k\to\infty} \wG_j(\r,\r';k) =0
\label{Gto0}
\ee
for any complex $k$ and $\r,\r'\in V_j$. In fact, the integral of $\wG_j(\r,\r';k')/(k-k')$ over the circumference of an infinitely large circle in the complex $k'$-plane (contour $C_1$ in \Fig{Fig:contour}) is zero, so using Cauchy's residue theorem and Lorentz reciprocity \cite{DoostPRA13} yields \Eq{GF-ML}, which in turn determines the RS normalization \cite{MuljarovEPL10,DoostPRA14,MuljarovPRB16Purcell,MuljarovOL18,SauvanOE22}.

However, \Eqs{F1}{F2} require that the coordinates of both GFs be in different regions, one inside and the other outside each resonator, in which case the ML series \Eq{GF-ML} becomes invalid that is equivalent to the fact that the RSs of a resonator are incomplete in the area outside it.
We solve this problem below by generalizing the ML series \Eq{GF-ML}, going beyond the resonator boundary.



Let one of the two coordinates of the GF $\wG_j$ be outside resonator $j$, namely, $\r'\notin V_j$ while $\r\in V_j$. Then \Eq{Gto0} is true only in the upper half of the complex $k$-plane, as follows from the outgoing boundary conditions of the GF. Integrating $\wG_j(\r,\r';k')/(k-k')$ over a closed contour, shown in \Fig{Fig:contour}, which consists of an infinite semicircle $C$ [again, with a vanishing result of integration due to \Eq{Gto0}] and a straight line $\Gamma$ \cite{footnote6}, we obtain a generalization of the ML series of the GF:
\bea
\wG_j(\r,\r';k)\!&=&\!\!\sum_n \frac{\wF_n^{(j)}(\r)\otimes \wF_n^{(j)}(\r')}{k-k_n^{(j)}}
-\frac{1}{2\pi i} \!\int_{\Gamma}\!\!\!dk' \frac{\wG_j(\r,\r';k')}{k-k'}
\nonumber
\\
&=&\!\sumint{n} \frac{\wF_n^{(j)}(\r)\otimes \wF_n^{(j)}(\r')}{k-k_n^{(j)}} \,,
\label{GF-MLc}
\eea
valid for $\r\in V_j$ and $\r'\notin V_j$. Here, we have used a remarkable property of the GF, proven in \cite{SI}, that if $\r$ and $\r'$ are in different regions, $\wG_j$ has a factorizable form,
\be
\wG_j(\r,\r';k')=-2\pi i \sum_s \wA^{(j)}_s(\r;k')\otimes \wB^{(j)}_s(\r';k')\,,
\label{AB}
\ee
where $\wA^{(j)}_s(\r;k')$ and $\wB^{(j)}_s(\r';k')$ are some vector fields on the contour $\Gamma$ which are added in \Eq{GF-MLc} to the set of the RSs $\wF_n^{(j)}(\r)$, and the index $s$ labels symmetry channels (or their combinations) of symmetric (or non-symmetric) resonators \cite{footnote4}. We also
adopted the notations from Refs.\,\cite{DoostPRA13,NealePRB20}, combining the RSs and the continuum of modes along $\Gamma$ into one set and writing the GF in the compact ML form \Eq{GF-MLc}, in which the integral stands for the modes of the continuum while the sum refers to the RSs above the contour $\Gamma$,
with the index $n$ now labeling all of these modes.


The ML expansion \Eq{GF-MLc}
is exactly what is needed for solving \Eqs{F1}{F2}. However, in dispersive systems, using the expansions \Eq{GF-MLc} as they are would result in a non-linear eigenvalue problem with respect to the eigenvalue $k$. To linearise it for the Drude-Lorentz dispersion \cite{SehmiPRB17} of $\wP_j(\r;k)$, we use alternative representations of the GF \cite{MuljarovPRB16,MuljarovOL18} along with \Eq{GF-MLc}, to finally obtain
\be
\wF(\r)=
\begin{cases}
\displaystyle
\sumint{n} c_n^{(1)} \wF_n^{(1)}(\r) \rmfor \r\in V_1\,,\\
\displaystyle
\sumint{m} c_m^{(2)} \wF_m^{(2)}(\r) \rmfor \r\in V_2\,.
\end{cases}
\label{Fsum}
\ee
Here, the expansions coefficients $c_n^{(1)}$ and $c_m^{(2)}$ of the field $\wF(\r)$ and the wave number $k$ of a RS of the full system are solutions of the {\it linear} eigenvalue problem represented by the following set of simultaneous equations, for all $n$ and $m$ included in \Eq{Fsum}:
\bea
(k-k_n^{(1)}) c_n^{(1)}&=&-k \sumint{m} U_{nm}^{(2)}(\infty) c_m^{(2)}
\nonumber\\
&&+ k_n^{(1)} \sumint{m} [U_{nm}^{(2)}(\infty)- U_{nm}^{(2)}(k_n^{(1)})] c_m^{(2)}\,,
\nonumber\\
(k-k_m^{(2)}) c_m^{(2)}&=&-k \sumint{n} U_{mn}^{(1)}(\infty) c_n^{(1)}
\label{CT}
\\
&&+ k_m^{(2)} \sumint{n} [U_{mn}^{(1)}(\infty)- U_{mn}^{(1)}(k_m^{(2)})] c_n^{(1)}\,,
\nonumber
\eea
where the matrix elements are given by
\bea
U_{mn}^{(1)}(q)&=&\int_{V_1} d\r  \wF_m^{(2)}(\r) \cdot [\wP_1(\r;q)-\wP_b] \wF_n^{(1)}(\r)\,,
\nonumber
\\
U_{nm}^{(2)}(q)&=&\int_{V_2} d\r  \wF_n^{(1)}(\r) \cdot [\wP_2(\r;q)-\wP_b] \wF_m^{(2)}(\r)\,,
\label{MatrElem}
\eea
see \cite{SI} for details of derivation.

\begin{figure}[t]
\includegraphics*[clip,width=0.6\textwidth]{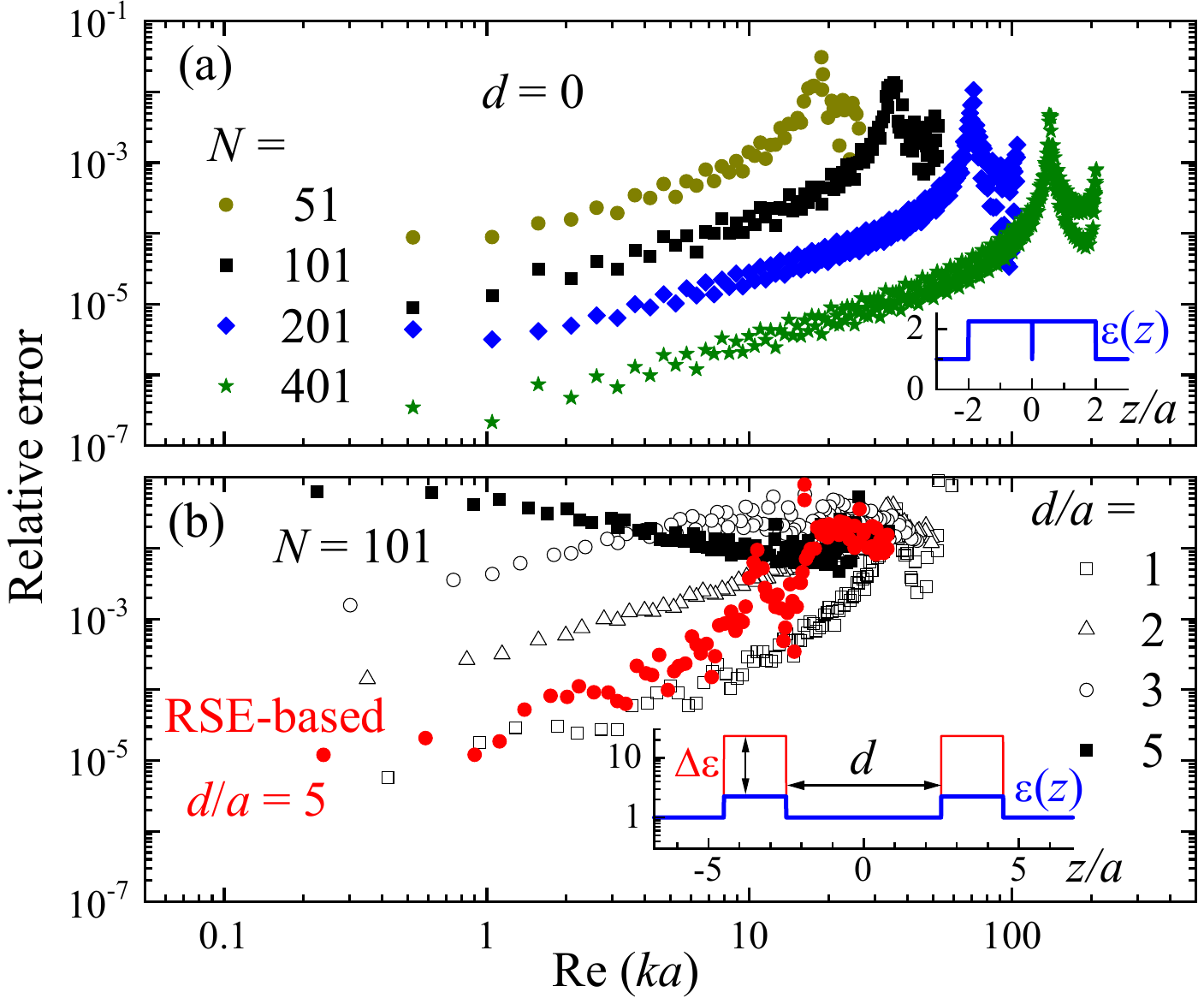}
\vspace{-10.0mm}
		\caption{
Relative error for the RS wave numbers of the coupled slab resonators, separated by distance $d$, calculated by the present theory \Eq{CT} with continuum-to-RS ratio $f=1.5$, (a) for $d=0$ and the number $N$ of basis RSs as given and (b) for $N=100$ and $d$ as given, demonstrating a fundamental limitation as $d$ increases. The error of the RSE-based approach for $d=5a$ and $N=100$ is shown by red circles in (b).
Insets: permittivity profiles of the coupled resonators and perturbation $\Delta\epsilon$ used in the RSE-based approach to \Eq{CT}.}
\label{Fig:error}
	\end{figure}

{\em Two slabs.} We verify the rigorous theory of coupled resonators presented above for the system of two identical dielectric slabs of width $2a$ each, separated by a distance $d$. For $d=0$, the eigenvalues $k$ of \Eq{CT} (red crosses in \Fig{Fig:theories}) are in agreement with the exact solution (blue squares), with the relative error [\Fig{Fig:error}(a)] reducing as $1/N^3$, similar to the RSE convergence \cite{MuljarovEPL10}. Here, $N$ and $fN$ are, respectively, the number of the RSs and discretized modes of the continuum of an individual resonator (single slab), taken into account in \Eq{CT} and shown in \Fig{Fig:theories} by black and gray stars. The discretization of the continuum has an effect on the accuracy similar to the truncations, in which only the RSs with $|k_n|<k_{\rm max}(N)$ are taken into account (as in the RSE), so $f$ should in principle also increase with $N$ but is kept fixed in this work at its optimal value $f=1.5$.

Increasing the distance $d$ has a dramatic effect on the error, as seen in \Fig{Fig:error}(b), in which  large errors (and no convergence) are seen already for $d/a=5$. This is a consequence of the exponential growth of the wave functions outside the resonator \cite{DoostPRA12,LalanneLPR25}, which is a severe limitation for using \Eq{CT} in practice. Moreover, this is a common problem for any resonator, as its spectrum contains an infinite countable number of Fabry-P\'erot (FP) RSs \cite{SztranyovszkyPRA22}, having
a rather large imaginary part of $k_n$, responsible for the exponential growth. This fundamental problem is solved below by combining the present approach with the RSE.



{\em RSE-based approach.} The RSE allows one to accurately and efficiently find the RSs of a target system using the RSs of a basis system \cite{MuljarovEPL10,DoostPRA12,DoostPRA13,DoostPRA14,MuljarovOL18}. In the present case, the target system is a single resonator with the average permittivity $\epsilon$, while the basis system can be chosen as a similar or simpler resonator with a much higher permittivity $\epsilon+\Delta\epsilon$, so its FP modes have sufficiently small imaginary part $\gamma_0\approx (a\Delta\epsilon)^{-1}$, where $2a$ is the shortest size of the basis system.
For error reduction, the contour $\Gamma$ (\Fig{Fig:contour}) should be chosen not too far away from the real axis and  not too close to the FP modes.
Low-Q modes, such as leaky modes of a spherical cavity \cite{SztranyovszkyPRA22}, can be left outside the contour $\Gamma$, as illustrated in \Fig{Fig:contour}. Therefore, the imaginary part of the continuum modes on the contour $\Gamma$  can be chosen as $\gamma=f_\gamma \gamma_0$. At the same time, the exponential growth of the continuum modes should be limited by a constant, $\gamma(d+2a)=f_d$, which results in a distant dependent permittivity perturbation $\Delta\epsilon=(d/a+2)f_\gamma/f_d$ of the RSE-based approach. The constants $f_\gamma$ and $f_d$ are therefore two parameters of the present theory having the optimal values for planar structures $f_\gamma\approx12$ and $f_\gamma/f_d\approx2$, see \cite{SI} for details.


We demonstrate this approach for planar coupled resonators separated by distances of up to $d/a=40$. For two glass slabs in vacuum, separated by $d/a=5$, the errors are shown by red stars in \Fig{Fig:error}(b), using $\Delta\epsilon=24$ as RSE perturbation, demonstrating a computational quality similar to the $d=0$ case without RSE, see \cite{SI} for more results.

The RSE plays another crucial role in the present approach: It allows us to rigorously prove the factorizable form of the GF \Eq{AB} and to determine the ML expansion \Eq{GF-MLc} of {\rm any} resonator.
In fact,
the GF of an arbitrary resonator is not known analytically, so there is generally no way to determine the continua of modes contributing to \Eq{GF-MLc}, and different symmetry channels in \Eq{AB} are mixed. However, transforming an analytically solvable resonator into an arbitrary resonator within the RSE framework preserves the validity of \Eq{GF-MLc} and determines all the modes contributing to it \cite{SI}.





\begin{figure}[t]
\includegraphics*[clip,width=0.6\textwidth]{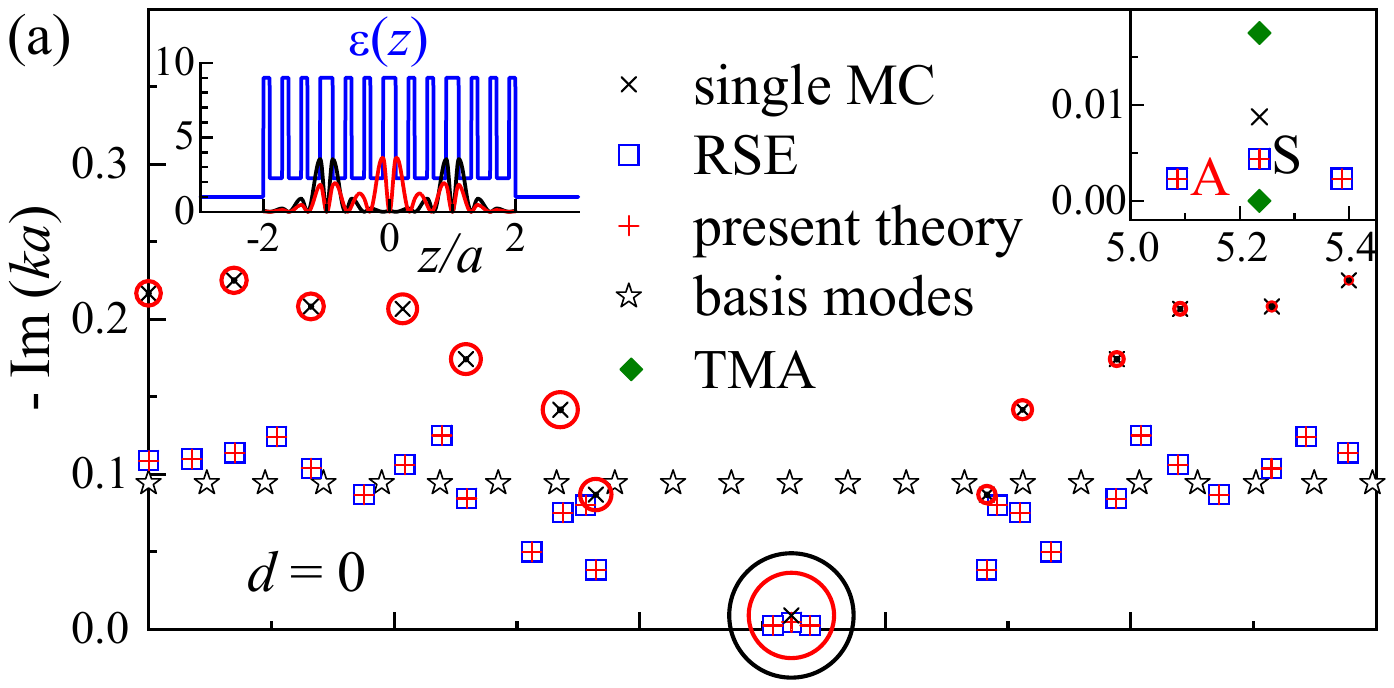}
\vskip-35mm
\includegraphics*[clip,width=0.6\textwidth]{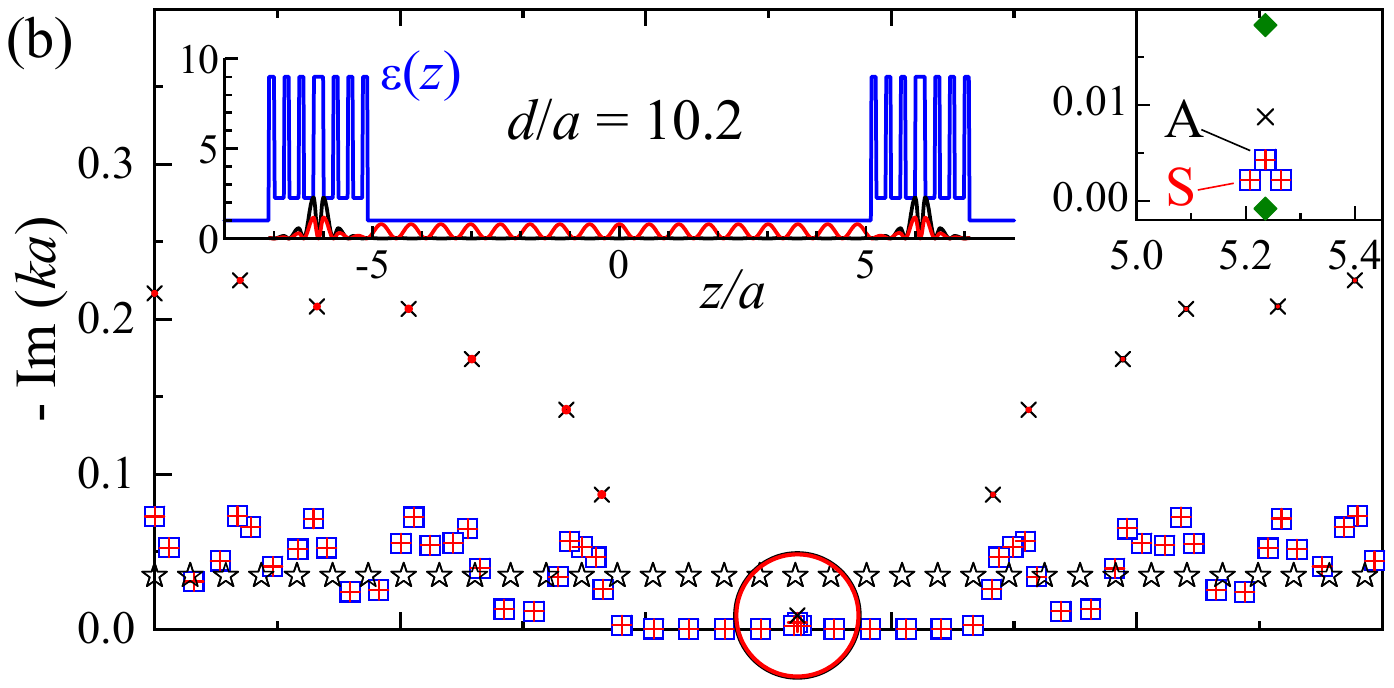}
\vskip-35mm
\includegraphics*[clip,width=0.6\textwidth]{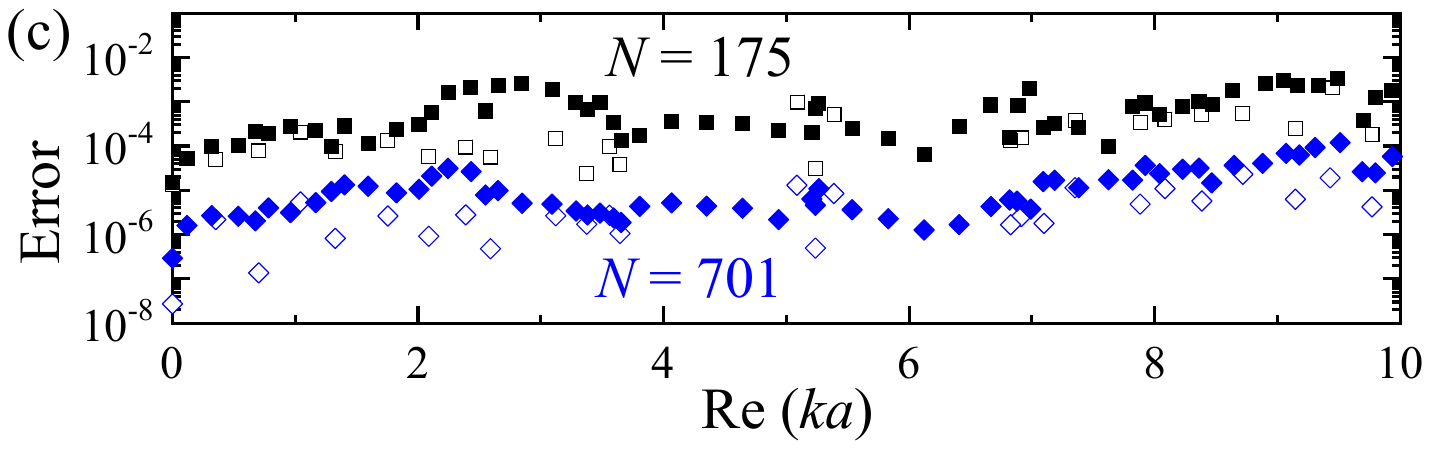}
\vspace{-54mm}
\caption{ RS wave numbers of coupled MCs, separated by distance (a)  $d=0$ and (b) $d=10.2a$, calculated by the RSE \cite{DoostPRA12} (open blue squares) and the present RSE-based theory (red crosses $+$), along with those of a single MC (black crosses x) and the basis slab (stars). The expansion coefficients $b_n$ at the single-MC RSs for RSs A and S of the coupled MCs in the right insets are given by the color-coded circles around the single-MC wave numbers, with the circle area proportional to $|b_n|^2$. Right insets: zoom-in of the spectra in (a) and (b) around the fundamental CM of the single MC and TMA (green diamonds), demonstrating the failure of the tight-binding concept.
Left insets: profiles of the permittivity (blue) of coupled MCs and the electric field $|E|^2$ of the modes S and A in the right insets (color coded).
(c) Relative error for the RSs wave numbers of the coupled MCs in (a) and (b), shown, respectively, by open and full black squares ($N=175$) and blue diamonds ($N=701$).
 }
\label{Fig:TwoMCs}
	\end{figure}

{\em Coupled MCs.} Applying the RSE-based theory to two $\lambda/2$ MCs, the cavity mode (CM) of a single MC splits by symmetry into 3 high-Q modes for $d=0$ [see \Fig{Fig:TwoMCs}(a) and right inset], and even more high-Q modes are formed between the MCs for $d=10.2a$ [\Fig{Fig:TwoMCs}(b)]. All the modes of the coupled system are well reproduced by the present theory (red crosses) within the shown spectral range, and the error scales as $1/N^3$ [\Fig{Fig:TwoMCs}(c)], see \cite{SI} for more details and other examples of coupled resonators.

Mapping the expansion \Eq{Fsum} back onto the RSs only (no continuum) within each MC where its RSs are  complete, $\wF(\r)=\sum_n b_n^{(j)} \wF_n^{(j)}(\r)$ for $\r\in V_j$, one can see the mode hybridization and the contribution of all the modes of the singe MC (black crosses x), shown by red and black circles (with the circle area proportional to $|b_n|^2$) for symmetric (S) and antisymmetric (A) coupled modes [right insets in \Fig{Fig:TwoMCs}(a),(b)], for which $b_n=b_n^{(1)}=\pm b_n^{(2)}$ by symmetry. In particular, mode S in \Fig{Fig:TwoMCs}(a) and both modes A and S in \Fig{Fig:TwoMCs}(b) have the dominant contributions of the single CMs [see also the wave functions in the left insets in \Fig{Fig:TwoMCs}(a),(b)], so the mode hybridization occurs as expected. However, using this information in the spirit of the tight-binding approach results in an entirely wrong and unphysical mode splitting as demonstrated by the TMA (green diamonds). Note that the TMA obtained from \Eqs{CT}{MatrElem} by keeping one CM per each resonator is identical to the standard tight-binding model.

\Fig{Fig:TwoMCs} demonstrates that the famous concept of tight binding fails in the case of open coupled resonators, even for high-Q modes \cite{footnote5}. As has been highlighted in Ref.\,\cite{LalanneLPR25}, owing to the exponential growth of the RSs outside a resonator these RSs increasingly perturb with distance, when two or more resonators are coupled, in disagreement with the intuition that distant objects do not interact. In the present case of the hybridized bonding and antibonding supermodes, dominated by the CMs, the seemingly small contribution of other RSs (with $|b_n|^2$ below 1\% compared to the CMs) cannot be neglected due to their exponential growth and consequently strong coupling between the resonators.

{\em Conclusion.} We have developed a rigourous theory of coupled resonators, based on a generalization of the Mittag-Leffler expansion of the dyadic Green's function beyond the resonator boundary, by adding a continuum of modes to the incomplete set of resonant states. To circumvent the fundamental limitation of the theory caused by the exponential growth of the resonant states outside a resonator, we have combined this theory with the resonant-state expansion. This provides also a rigorous proof of the formalism and a reliable calculation of the continuum, which are otherwise not possible, since the Green's function of an  arbitrary system is not known analytically. We have also shown that the concept of tight binding fails in the case of open resonators. This is in a drastic contrast with recent claims and illustrations \cite{VialJO16,LiuPRB20,CogneePhD20,HughesPRX20,BachelardOE22}.


\begin{thebibliography}{43}
\bibitem [{foo({\natexlab{b}})}]{footnote2}%
  \BibitemOpen
  \href@noop {} {\bibinfo {title} {This is a harder case to address from the
  viewpoint of completeness, since otherwise the RSs of both systems would be
  complete in the overlap region without adding continua.}}
\bibitem [{foo({\natexlab{c}})}]{footnote6}%
  \BibitemOpen
  \href@noop {} {\bibinfo {title} {A similar integration, though over a finite
  closed contour is used in the Riezs-projection theory \cite{BurgerPRA18}.}}
\bibitem [{foo({\natexlab{d}})}]{footnote4}%
  \BibitemOpen
  \href@noop {} {\bibinfo {title} {For example, $s$ is the orbital quantum number for a spherically symmetric resonator and the parity for a planar system with mirror symmetry, although even- and odd-parity states of the continuum can always  be combined into one channel.}}
\bibitem [{foo({\natexlab{e}})}]{footnote5}%
  \BibitemOpen
  \href@noop {} {\bibinfo {title} {This concept may still be working for a
  subspace of infinite-Q modes, such as bound states in the continuum in
  coupled photonic-crystal resonators, as recently demonstrated in Ref.
  \cite{CanosPRL25}.}}
\end{thebibliography}
\end{document}